\documentclass{article}
\usepackage[text={6in,9in},centering]{geometry}

\usepackage{authblk}

\usepackage{cite}

\usepackage[braket, qm]{qcircuit}

\usepackage{amsmath}

\usepackage{graphicx}

\usepackage{here}

\makeatletter
\DeclareRobustCommand
  \myvdots{\vbox{\baselineskip4\p@ \lineskiplimit\z@
    \hbox{.}\hbox{.}\hbox{.}}}
\makeatother

\title{Estimation of the number of logical qubits required for radiation transport calculations with a quantum computer}
\author{Takuma Noto \thanks{Email: t.noto@shimz.co.jp}}
\affil{Institute of Technology, Shimizu Corporation, 3-4-17, Etchujima, Koto-ku, Tokyo 135-8530, Japan}
\date{\today}

\begin{document}

\maketitle

\begin{abstract}
    As an application of fault-tolerant quantum computers, we consider radiation transport calculations in this study.
    Radiation transport calculation using Monte Carlo calculation can obtain a solution to even a problem difficult to solve analytically.
    However, it is time-consuming depending on the scale and precision of the problem.
    Because it is known that the computational complexity of Monte Carlo calculation can be square rooted by quantum amplitude estimation, acceleration can be expected if radiation transport calculation is run on a quantum computer.
    In this study, we designed and investigated a quantum circuit for a simplified transport calculation in which the reaction is only forward scattering or absorption and the energy and time do not change as well as showed the possibility of acceleration the calculation.
    Further, we estimated the number of logical qubits required to solve practical problems based on the quantum circuit.
\end{abstract}


\section{Introduction}

Monte Carlo (MC) calculation is a general term for methods that solve problems using random numbers. In the 1940s, von Neumann and Ulam proposed a method for statistically simulating the behavior of neutrons, and the actual calculation by the world's first computer ENIAC was the beginning of MC calculation using a computer \cite{1}. It has become possible to find the solution to a problem that is difficult to solve analytically via simulation. At present, the behavior of radiation including neutrons can be calculated accurately, even with complicated geometries, by the latest three-dimensional (3D) MC radiation transport calculation codes, such as MCNP6 \cite{2}, PHITS3 \cite{3}, and GEANT4 \cite{4}. MC calculations are employed in various fields, such as nuclear reactor and accelerator design, medical application, detector characteristic investigation, and space field. However, to handle the enormous size and high precision of some problems, MC calculations require numerous trials and are time-consuming, even if the calculation efficiency is improved by the dispersion reduction method. Because the order of statistical error in $N$ evaluation of classical trials using random numbers is $N^{-1/2}$, the number of trials required to reduce the error to 1/10 is 100 times larger.

Moreover, quantum algorithms using quantum behavior such as superposition and entanglement have been proposed to reduce the computational complexity significantly compared with the classical algorithms. Accelerating MC calculation \cite{5} by quantum amplitude estimation (QAE) \cite{6,7,8} is one of such algorithms and the order of error is $N^{-1}$ after evaluating the quantum trials $N$ times. Therefore, compared with the classical MC calculation, the computational complexity can be reduced in the order of the square root, which is expected to accelerate the calculation.

In this study, we investigate radiation transport calculation as one of the applications of fault-tolerant quantum computers. Particularly, as a basic study for running radiation transport calculation on a quantum computer, we design a quantum circuit for simplified radiation transport calculation of virtual particles, examined the possibility of accelerating the calculation, and estimated the number of logical qubits required for practical applications.

\section{Quantum Circuit for a Simplified Radiation Transport Calculation}

\subsection{Simplified radiation transportation}

Figure 1 shows a classic flowchart of a simplified radiation transport calculation. Radiation has both particle and wave properties; however, it is considered a particle in MC calculations. To simplify the transport process, the changes in direction, energy, and time in the transport as well as the generation of secondary particles are omitted in the calculation. Therefore, the particle behavior is only repeated forward scattering or stopping due to absorption reaction. In addition, the position and amount of movement that particles can take are only nonnegative integers. The dispersion reduction method is not used. The calculation procedure in Figure 1 is as follows: (1) the determination of the source position of the particle, (2) the determination of the flight distance depending on the region where the particle is located, (3) update to the particle position after the flight, and (4) the determination of the reaction type of particle and material in the region. If the reaction is scattering, repeat steps (2)–(4), whereas if it is absorption, return to (1) to finish the calculation of the current particles and move to transport new particles. In a typical radiation transport calculation, the range of space and energy to be handled in the calculation is set, and the calculation is terminated when the particles exceed that range. However, in the simplified radiation transport calculation, the upper limit of the number of flights $n$ is set as the termination condition of the calculation. The flight distance $d$ of a particle has a distribution that can be calculated by $d=-\lambda \ln \eta $ from the mean free path $\lambda$ of the particle in the material and the uniform random number $\eta$ ($0 \leq \eta \leq 1$). If the updated position of the particle crosses the boundary with the adjacent region, the particle is not stopped at the boundary. The reaction is determined using a random number from the cumulative distribution function of the reaction probability. As a result, dividing the number of particles that have reached an arbitrary position by the total number of trials gives the probability that the particles will reach that position.

\begin{figure}[h]
\centering
\includegraphics[width=84truemm]{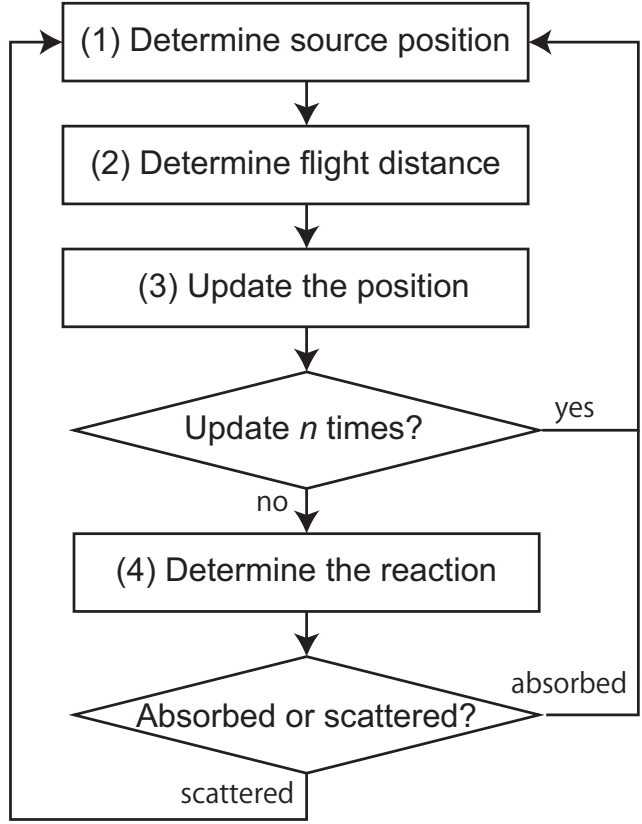}
\caption{Flowchart of classic simplified radiation transport calculation}
\end{figure}

\subsection{Quantum circuit}

Figure 2 shows a quantum circuit that realizes the simplified radiation transport calculation. The circuit implemented qubits with a superimposed probability distribution of the flight distance and reaction type instead of classical random sampling. Register for $x$ position ($Reg. X$) is a quantum register that records the latest position of particles. Register for flight distance ($Reg. D_m$) and register for reaction ($Reg. R_m$) are quantum registers that record the flight distance and reaction in the $m$-th flight, respectively. Ancilla for progress ($Anc. P$) is an ancillary qubit that determines whether the position update is effective from the past reaction history. Setting the initial value of qubits to \ket{0} corresponds to (1) of the classic algorithm. Below are descriptions of operations in the quantum circuit.

Operation $\mathcal{P}$: $\mathcal{P}$ is an operation that implements the superposition state of the probability distribution of flight distance and reaction type in $Reg. D_m$ and $Reg. R_m$, respectively, based on where the particle is located. By defining \ket{0} and \ket{1} as absorption and scattering reactions, respectively, the state \ket{r_m} of the $m$-th reaction rotated by $\mathcal{P}$ store a proper superposition of \ket{0} and \ket{1}. The quantum state \ket{d_m} of the $m$-th flight distance is a superposition state of \ket{0} to $\ket{d_{\max}}$, where $d_{\max}$ (a positive integer) is the maximum flight distance. This operation corresponds to (2)–(4) of the classical algorithm.

Multicontrolled Toffoli gate (MCT): MCT store \ket{1} in $Anc. P$ only when all past reactions are scattered. On the $n$-th flight, an $n$-controlled Toffoli gate is required to confirm that all of $Reg. R_1$ to $Reg. R_n$ are \ket{1}. This operation is peculiar to the quantum algorithm because the classical algorithm moves to the calculation of the next particle when the particle is absorbed.

Controlled $\varSigma$: An operation to add the flight distance of $Reg. D_m$ to the position $Reg. X$ when $Anc. P$ is \ket{1}. This operation corresponds to a controlled quantum adder and (3) of the classical algorithm.

The state \ket{x_n} of particle positions after flighting $n$ times are stored in $Reg. X$ by repeating the circuits. A two-region three-flight quantum circuit based on this algorithm is shown in Appendix.

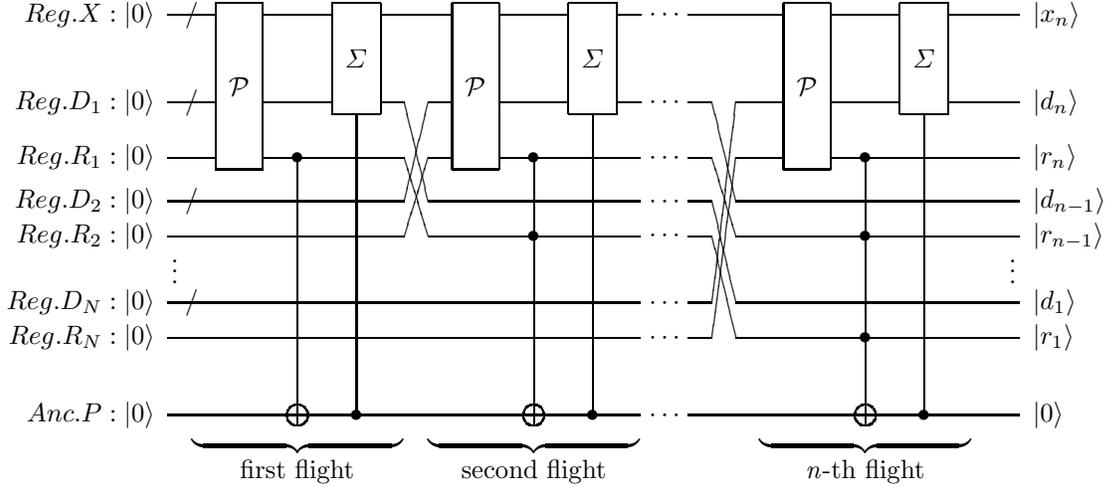
\begin{figure}[h]
\[\Qcircuit @C=0.9em @R=1.2em {
  & \lstick{Reg.X   : \ket{0} }  & \qw {/}           & \multigate{3}{\mathcal{P}} & \qw      & \multigate{2}{\varSigma } & \qw & \qw            & \multigate{3}{\mathcal{P}} & \qw      & \multigate{2}{\varSigma } & \qw & \cdots & & \qw & \qw            & \qw & \multigate{3}{\mathcal{P}} & \qw      & \multigate{2}{\varSigma } & \qw & \qw               & \rstick{\ket{x_n} }     \qw \\
  &                              &                   &                  &          &                           &     &                &                  &          &                           &     &        & &     &                &     &                  &          &                           &     &                   &                             \\
  & \lstick{Reg.D_1 : \ket{0} }  & \qw {/}           & \ghost{P}        & \qw      & \ghost{\varSigma }        & \qw & \link{2}{-1}   & \ghost{P}        & \qw      & \ghost{\varSigma }        & \qw & \cdots & & \qw & \link{5}{-1}   & \qw & \ghost{P}        & \qw      & \ghost{\varSigma }        & \qw & \qw               & \rstick{\ket{d_n} }     \qw \\
  & \lstick{Reg.R_1 : \ket{0} }  & \qw               & \ghost{P}        & \ctrl{7} & \qw                       & \qw & \link{2}{-1}   & \ghost{P}        & \ctrl{7} & \qw                       & \qw & \cdots & & \qw & \link{5}{-1}   & \qw & \ghost{P}        & \ctrl{7} & \qw                       & \qw & \qw               & \rstick{\ket{r_n} }     \qw \\
  & \lstick{Reg.D_2 : \ket{0} }  & \qw {/}           & \qw              & \qw      & \qw                       & \qw & \link{-2}{-1}  & \qw              & \qw      & \qw                       & \qw & \cdots & & \qw & \link{-2}{-1}  & \qw & \qw              & \qw      & \qw                       & \qw & \qw               & \rstick{\ket{d_{n-1}} } \qw \\
  & \lstick{Reg.R_2 : \ket{0} }  & \qw               & \qw              & \qw      & \qw                       & \qw & \link{-2}{-1}  & \qw              & \ctrl{0} & \qw                       & \qw & \cdots & & \qw & \link{-2}{-1}  & \qw & \qw              & \ctrl{0} & \qw                       & \qw & \qw               & \rstick{\ket{r_{n-1}} } \qw \\
  &                              & \lstick{\myvdots} &                  &          &                           &     &                &                  &          &                           &     &        & &     &                &     &                  &          &                           &     & \rstick{\myvdots} &                             \\
  & \lstick{Reg.D_N : \ket{0} }  & \qw {/}           & \qw              & \qw      & \qw                       & \qw & \qw            & \qw              & \qw      & \qw                       & \qw & \cdots & & \qw & \link{-3}{-1}  & \qw & \qw              & \qw      & \qw                       & \qw & \qw               & \rstick{\ket{d_1} }     \qw \\
  & \lstick{Reg.R_N : \ket{0} }  & \qw               & \qw              & \qw      & \qw                       & \qw & \qw            & \qw              & \qw      & \qw                       & \qw & \cdots & & \qw & \link{-3}{-1}  & \qw & \qw              & \ctrl{0} & \qw                       & \qw & \qw               & \rstick{\ket{r_1} }     \qw \\
  &                              &                   &                  &          &                           &     &                &                  &          &                           &     &        & &     &                &     &                  &          &                           &     &                   &                             \\
  & \lstick{Anc.P   : \ket{0} }  & \qw               & \qw              & \targ    & \ctrl{-8}                 & \qw & \qw            & \qw              & \targ    & \ctrl{-8}                 & \qw & \cdots & & \qw & \qw            & \qw & \qw              & \targ    & \ctrl{-8}                 & \qw & \qw               & \rstick{\ket{0} }       \qw \\
  &                              &                   &                  & \underbrace{\hspace{8em}}_{\mbox{first flight}} & & &         &                  & \underbrace{\hspace{8em}}_{\mbox{second flight}} & & & & &     &                &     &                  & \underbrace{\hspace{8em}}_{\mbox{\textit{n}-th flight}} & & & \\
  }\]
\caption{Quantum circuit for simplified radiation transport calculation}
\end{figure}

\subsection{Position readout by QAE}
It is possible to estimate quantum amplitude $p$ by QAE when the interesting state \ket{\varPsi_1} whose amplitude is $p$ and the other state \ket{\varPsi_0} are created by unitary operation $\mathcal{A}$ as follows:
\begin{equation}
\mathcal{A} \ket{0} = \sqrt{1-p} \ket{\varPsi_0} + \sqrt{p} \ket{\varPsi_1}
\end{equation}

Letting the state after a maximum of $n$ flights in the transport calculation circuit represent \ket{\varPhi}\ket{x_n} and the series of operations for $\mathcal{F}$, the circuit can be expressed as follows:
\begin{equation}
\mathcal{F} \ket{0} \ket{0} = \ket{\varPhi} \ket{x_n}
\end{equation}

Here, one ancillary qubit is added to the circuit. An operation $\mathcal{R}$ store \ket{1} in the ancillary qubits when \ket{x_n} is in a specific state. In this case, a series of operations comprising $\mathcal{F}$ and $\mathcal{R}$ corresponds to $\mathcal{A}$, and the $p$ is the quantum amplitude that \ket{x_n} is in a specific state.
\begin{equation}
\mathcal{R} \mathcal{F} \ket{0} \ket{0} \ket{0} = \mathcal{A} \ket{0} \ket{0} \ket{0} = \ket{\varPhi} \ket{x_n}(\sqrt{1-p} \ket{\varPsi_0} + \sqrt{p} \ket{\varPsi_1})
\end{equation}

Therefore, by the QAE of this ancillary qubit, the probability that the particle will be located at a specific position after flights can be calculated. As an example, the probability that \ket{x_n} is greater than or equal to $2^k$ ($k$ is a nonnegative integer) after the flights can be obtained by rotating the ancillary qubit to \ket{1} when the qubits of $Reg. X$ except for the lower $k$ qubits contains \ket{1}.

\section{Discussion and Conclusion}
A quantum circuit for one-dimensional radiation transport calculation was designed. The probability distribution of the position is completely stored in $Reg. X$ by one quantum trial $\mathcal{F}$. However, it is necessary to use QAE to estimate the probability. The error of estimation after repeating the operation including $\mathcal{F}$ for $N$ times is on the order of $N^{-1}$. In the classical algorithm, most particles will never have a maximum number of flights $n$ because the calculations stop when radiation deviates from the set conditions. If only absorption and scattering occur in a single region and the probability of absorption in one reaction is $p_a$, the expected value $n_E$ of the number of flights can be calculated as follows:
\begin{equation}
n_E = \displaystyle \sum_{k}^{\infty}k p_a(1-p_a)^{k-1}=\frac{1}{p_a}
\end{equation}

Assuming $n$ is the linear estimate of $n_E$, the order of computational complexity of quantum and classical trials is the same. Because the order of the error is $N^{-1/2}$ in the classical MC calculation when the number of trials is $N$, it is possible to reduce the amount of calculation by the order of the square root using the quantum algorithm.
Moreover, the values of flight distance and reaction type recorded in memory can be overwritten in the classical MC calculation, so that the number of trials can be increased even with small memory. However, on the quantum circuit, the position register after the flight is in a state of quantum entanglement with each reaction and flight distance, so quantum registers that hold the reaction types and flight distances are necessary. Therefore, the required number of qubits increases linearly with the maximum number $n$ of flights. To extend the circuit to practical radiation transport calculations, $n$ quantum registers are required for each of at least 7 changes: the 3D position, 3D direction of the motion vector, and energy. Because one set of registers to store the final value is also required, $7(n + 1)$ quantum registers are required to store the final and history states. In addition, because as many quantum adders are required to update states, $7n$ quantum adders are needed. Assuming that each variable is considered a 32-bit floating-point number, it has been reported that the number of qubits required for a quantum adder is 140 [9], which is two 32-qubit variables and 76 ancillary qubits. Further, $n$ qubits to record $n$ reactions and an ancillary qubit to check the reaction history are required. Therefore, it is estimated that at least $32 \times 7 \times (n + 1) + 76 \times 7 \times n + n + 1 = 757 \times n + 225$ qubits are required. Although QAE is used to estimate the probability distribution, it is possible to estimate specific amplitudes by adding one qubit to specify the state by hybrid calculation with a classical computer [8].

Further, as a practical circuit, we assumed a problem that the radiation penetrates a region 100 times the mean free path. Assuming the maximum number of flights is $n$ = 100, the circuit requires approximately $1 \times 10^5$ logical qubits. In addition, it is presumed that several orders of magnitude more will be needed to represent complex geometric systems, read nuclear date libraries containing data on reactions caused by interactions between particles and atoms, create secondary particles, and add angular dependence of change degree. We will study algorithms for more complicated transport calculations and study the performance from the viewpoint of circuit depth.

\section*{Statements and Declarations}
The author have no conflicts of interest to declare.

\setcounter{figure}{0}
\renewcommand{\thetable}{A\arabic{table}}
\renewcommand{\thefigure}{A\arabic{figure}}
\section*{Appendix: Two-region three-flight transport circuit}
We show a circuit for radiation transport calculation incident on a system comprising Regions 1 and 2, which are located at $0 \leq x < 4$ and $x > 4$, respectively, where $x$ is the radiation position. In the simplified radiation transport calculation of this study, the energy of particles does not change during transport and the scattering angle is always $0^\circ$. Table A1 shows the probability distribution of flight distance. The mean free paths in Regions 1 and 2 were set to 1.5 and 1.0, respectively. The probability distribution was calculated in increments of 0.1, and the flight distance was rounded to an integer value. Reaction probabilities in Regions 1 and 2 were, respectively, set to 0.25 and 0.40 for absorption and 0.75 and 0.60 for scattering. The maximum number of flights was 3. Because the maximum value of flight distance was 3 and only two types of reaction were handled, $Reg. D_m$ and $Reg. R_m$, which, respectively, hold the flight distance and reaction type in the $m$-th flight, are 2 qubits and 1 qubit, respectively. However, $Reg. R_1$ is omitted because it is always \ket{1}. $Reg. X$ needs 4 qubits because the maximum value of $x$ is 8. In addition, 1 qubit is required for $Anc. P$, and 1 ancillary qubit for the region ($Anc. R$) is added to determine the region where the particles are located after the flight.

By applying the quantum circuit shown in Figure A1 to the qubits for the position initialized at \ket{0}, a quantum state \ket{x_n} of the probability distribution of the position after a maximum of $n$ flights can be obtained. Below are definitions for operations in the circuit.

Operation $\mathcal{L}$: $\mathcal{L}$ is the operation to set $Anc. R$ to \ket{1} when the particle is in Region 2 or the particle position is 4 or more. Assuming the quantum register \ket{x} comprises multiple qubits and \ket{x_{[i]}} indicates the state of the $i + 1$th qubit of the register, the state of $\ket{x}=\ket{x_{[3]}} \otimes \ket{x_{[2]}} \otimes \ket{x_{[1]}} \otimes \ket{x_{[0]}}$ represents a positive integer $x = 2^3x_{[3]} + 2^2x_{[2]} + 2^1x_{[1]} + 2^0x_{[0]}$. $\mathcal{L}$ comprises a CNOT gate and a Toffoli gate to $Reg. X_{[2]}$, $Reg. X_{[3]}$, which are the two upper qubits of $Reg. X$, and $Anc. R$ (Figure A2).

Operation $\mathcal{Q}$: $\mathcal{Q}$ mainly comprises $y$-rotating gate $R_y$ (Figures A3 and A4) and stores flight distance and reaction probability distribution to $Reg. D_m$ and $Reg. R_m$, respectively, according to the state of $Anc. R$. The unitary matrix of the $R_y$ gate is as follows:
$$
R_y(\theta) = 
\begin{pmatrix}
\cos(\theta/2) & -\sin(\theta/2) \\
\sin(\theta/2) & \cos(\theta/2) \\
\end{pmatrix}
$$

The angle of rotation in Figure A3 is $\theta_1=2 \cos^{-1} \sqrt{0.25}$, $\theta_2=2 \cos^{-1} \sqrt{0.4}$. Because the first flight is always performed according to the classical calculation and $Reg. R_1$ is always \ket{1}, $Reg. R_1$ is omitted from the circuit. Probability distributions $P_{R1}(d)$ and $P_{R2}(d)$ have probabilities for the flight distance $d$ in Regions 1 and 2, respectively. The rotation angles $\theta_3$, $\theta_4$, and $\theta_5$ when the probability distribution $P(d)$ is input as the argument of the operation $R_D$ in Figure A3 are as follows:
$$
\theta_3 = 2 \cos^{-1} \sqrt{P(0) + P(1)}
$$

$$
\theta_4 = 2 \cos^{-1} \sqrt{\frac{P(2)}{P(2) + P(3)}}
$$

$$
\theta_5 = 2 \cos^{-1} \sqrt{\frac{P(0)}{P(0) + P(1)}}
$$

Operation $\mathcal{L}^{-1}$: This is an operation to initialize $Anc. R$ to \ket{0}.

MCT: Because the position is always updated after the first flight, it is omitted as in $Reg. R_1$. Therefore, CNOT is applied for the second flight and Toffoli gate is applied for the third flight.

Operation $\varSigma$: Some types of quantum adder have been reported. A quantum adder using \cite{10} can implement an adder without using an auxiliary qubit.

A series of operation $\mathcal{L}^{-1} \mathcal{Q} \mathcal{L}$ corresponds to operation $\mathcal{R}$ in Figure 2. The above quantum circuit was implemented using Qiskit \cite{11,12} and executed by its simulator. The probability distribution of the position obtained from the quantum amplitude of the quantum register \ket{x_n} is shown in Figure A5, and it is confirmed that it matches the one from the classical calculation. Further, because all operations are unitary, the probability that a particle is in a specific position can be estimated by QAE. As an example, if we want to see the probability that the particle is finally in Region 2, we can get the desired probability by applying the operation $\mathcal{L}$ to an ancillary qubit and performing QEA for the state where the ancillary qubit is \ket{1}.

\begin{table}[h]
\caption{Probability distribution of flight distance and reaction type in Region 1 and 2}
 \centering
  \begin{tabular}{c|cccc|cc}
                        & \multicolumn{4}{c|}{Flight distance} & \multicolumn{2}{c}{Reaction} \\
			& 0	& 1	& 2	& 3	& absorbed & scatter \\
    \hline
    Region 1	& 0.3	& 0.4	& 0.2	& 0.1	& 0.25	 & 0.75	\\
    Region 2	& 0.4	& 0.4	& 0.2	& 0	& 0.40	 & 0.60	\\
  \end{tabular}
\end{table}

\begin{figure}[H]
\centering
\[\Qcircuit @C=0.5em @R=1.2em {
  & \lstick{Reg.X   : \ket{0} }  & \qw {/_4} & \multigate{1}{\mathcal{L}} & \qw              & \multigate{1}{\mathcal{L}^{-1}} & \qw & \link{1}{-1}  & \qw                       & \qw & \link{1}{-1}   & \qw & \multigate{1}{\mathcal{L}} & \qw              & \multigate{1}{\mathcal{L}^{-1}} & \qw & \link{1}{-1}  & \qw                       & \qw & \link{1}{-1}   & \qw & \multigate{1}{\mathcal{L}} & \qw              & \multigate{1}{\mathcal{L}^{-1}} & \qw & \link{1}{-1}  & \qw                       & \qw & \rstick{\ket{x_3} }  \qw \\
  & \lstick{Anc.R   : \ket{0} }  & \qw       & \ghost{F}        & \multigate{2}{\mathcal{Q}} & \ghost{F^{-1}}        & \qw & \link{-1}{-1} & \multigate{2}{\varSigma } & \qw & \link{-1}{-1}  & \qw & \ghost{F}        & \multigate{3}{\mathcal{Q}} & \ghost{F^{-1}}        & \qw & \link{-1}{-1} & \multigate{2}{\varSigma } & \qw & \link{-1}{-1}  & \qw & \ghost{L}        & \multigate{3}{\mathcal{Q}} & \ghost{L^{-1}}        & \qw & \link{-1}{-1} & \multigate{2}{\varSigma } & \qw & \rstick{\ket{0} }    \qw \\
  &                              &           &                  &                  &                       &     &               &                           &     &                &     &                  &                  &                       &     &               &                           &     &                &     &                  &                  &                       &     &               &                           &     &                          \\
  & \lstick{Reg.D_1 : \ket{0} }  & \qw {/_2} & \qw              & \ghost{Q}        & \qw                   & \qw & \qw           & \ghost{\varSigma }        & \qw & \link{1}{-1}   & \qw & \qw              & \ghost{Q}        & \qw                   & \qw & \qw           & \ghost{\varSigma }        & \qw & \link{3}{-1}   & \qw & \qw              & \ghost{Q}        & \qw                   & \qw & \qw           & \ghost{\varSigma }        & \qw & \rstick{\ket{d_3} } \qw \\
  & \lstick{Reg.D_2 : \ket{0} }  & \qw {/_2} & \qw              & \qw              & \qw                   & \qw & \qw           & \qw                       & \qw & \link{1}{-1}   & \qw & \qw              & \ghost{Q}        & \ctrl{5}              & \qw & \qw           & \qw                       & \qw & \link{3}{-1}   & \qw & \qw              & \ghost{Q}        & \ctrl{5}              & \qw & \qw           & \qw                       & \qw & \rstick{\ket{r_3} } \qw \\
  & \lstick{Reg.R_2 : \ket{0} }  & \qw       & \qw              & \qw              & \qw                   & \qw & \qw           & \qw                       & \qw & \link{-2}{-1}  & \qw & \qw              & \qw              & \qw                   & \qw & \qw           & \qw                       & \qw & \link{-2}{-1}  & \qw & \qw              & \qw              & \qw                   & \qw & \qw           & \qw                       & \qw & \rstick{\ket{d_2} } \qw \\
  & \lstick{Reg.D_3 : \ket{0} }  & \qw {/_2} & \qw              & \qw              & \qw                   & \qw & \qw           & \qw                       & \qw & \qw            & \qw & \qw              & \qw              & \qw                   & \qw & \qw           & \qw                       & \qw & \link{-2}{-1}  & \qw & \qw              & \qw              & \ctrl{0}              & \qw & \qw           & \qw                       & \qw & \rstick{\ket{r_2} } \qw \\
  & \lstick{Reg.R_3 : \ket{0} }  & \qw       & \qw              & \qw              & \qw                   & \qw & \qw           & \qw                       & \qw & \qw            & \qw & \qw              & \qw              & \qw                   & \qw & \qw           & \qw                       & \qw & \link{-2}{-1}  & \qw & \qw              & \qw              & \qw                   & \qw & \qw           & \qw                       & \qw & \rstick{\ket{d_1} } \qw \\
  &                              &           &                  &                  &                       &     &               &                           &     &                &     &                  &                  &                       &     &               &                           &     &                &     &                  &                  &                       &     &               &                           &     &                          \\
  & \lstick{Anc.P   : \ket{0} }  & \qw       & \qw              & \qw              & \qw                   & \qw & \qw           & \qw                       & \qw & \qw            & \qw & \qw              & \qw              & \targ                 & \qw & \qw           & \ctrl{-6}                 & \qw & \qw            & \qw & \qw              & \qw              & \targ                 & \qw & \qw           & \ctrl{-6}                 & \qw & \rstick{\ket{0} }   \qw \\
  &                              &           &                  &                  & \underbrace{\hspace{10em}}_{\mbox{first flight}} & & &                    &     &                &     &                  &                  & \underbrace{\hspace{10em}}_{\mbox{second flight}} & & &                    &     &                &     &                  &                  & \underbrace{\hspace{10em}}_{\mbox{third flight}} & & & \\
  }\]
\caption{Quantum circuit for two-region three-flight radiation transport calculation}
\end{figure}
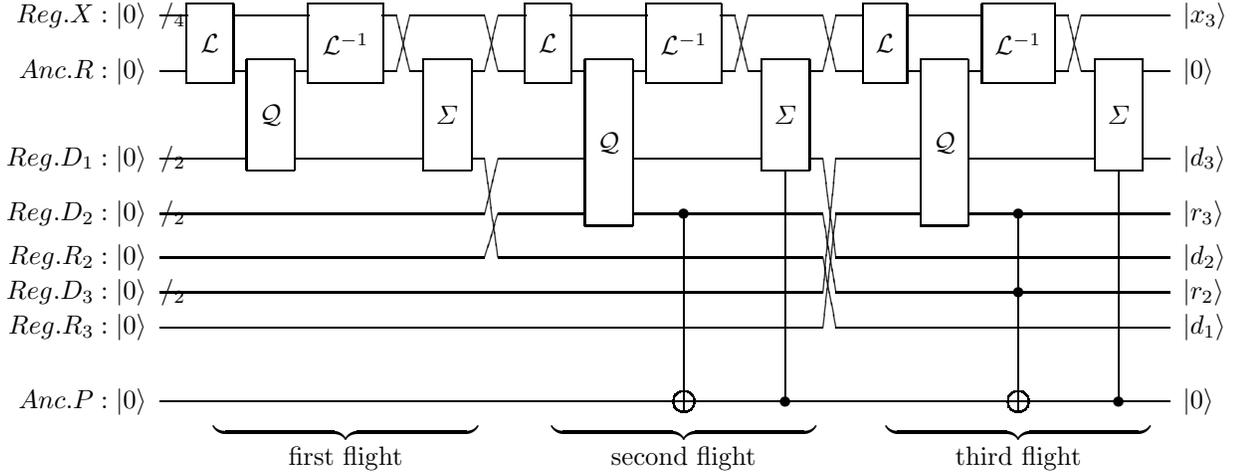

\begin{figure}[H]
\[\Qcircuit @C=1em @R=1.2em {
  & \lstick{Reg.X_{[3]}}     & \qw  & \ctrl{2} & \qw      & \ctrl{2} & \qw \\
  & \lstick{Reg.X_{[2]}}     & \qw  & \qw      & \ctrl{1} & \ctrl{0} & \qw \\
  & \lstick{Anc.R: \ket{0} } & \qw  & \targ    & \targ    & \targ    & \qw \\
  }\]
\caption{Quantum circuit for operation $\mathcal{L}$}
\end{figure}

\begin{figure}[H]
\[\Qcircuit @C=1em @R=1.2em {
  & \lstick{Anc.R}             & \qw       & \ctrl{1}           & \ctrl{2}             & \gate{X} & \ctrl{1}           & \ctrl{2}             & \gate{X} & \qw                     \\
  & \lstick{Reg.D_m: \ket{0} } & \qw {/_2} & \gate{R_D(P_{R1})} & \qw                  & \qw      & \gate{R_D(P_{R0})} & \qw                  & \qw      & \rstick{\ket{d_m} } \qw \\
  & \lstick{Reg.R_m: \ket{0} } & \qw       & \qw                & \gate{R_y(\theta_1)} & \qw      & \qw                & \gate{R_y(\theta_2)} & \qw      & \rstick{\ket{r_m} } \qw \\
  }\]
\caption{Quantum circuit for operation $\mathcal{Q}$}
\end{figure}

\begin{figure}[H]
\[\Qcircuit @C=1em @R=1.2em {
  & \lstick{Reg.D_{m[1]}} & \qw  & \gate{R_y(\theta_3)} & \ctrl{1}             & \gate{X} & \ctrl{1}             & \gate{X} & \qw \\
  & \lstick{Reg.D_{m[0]}} & \qw  & \qw                  & \gate{R_y(\theta_4)} & \qw      & \gate{R_y(\theta_5)} & \qw      & \qw \\
  }\]
\caption{Quantum circuit for operation $\mathcal{R}_D(P)$}
\end{figure}

\begin{figure}[H]
\centering
\includegraphics[width=84truemm]{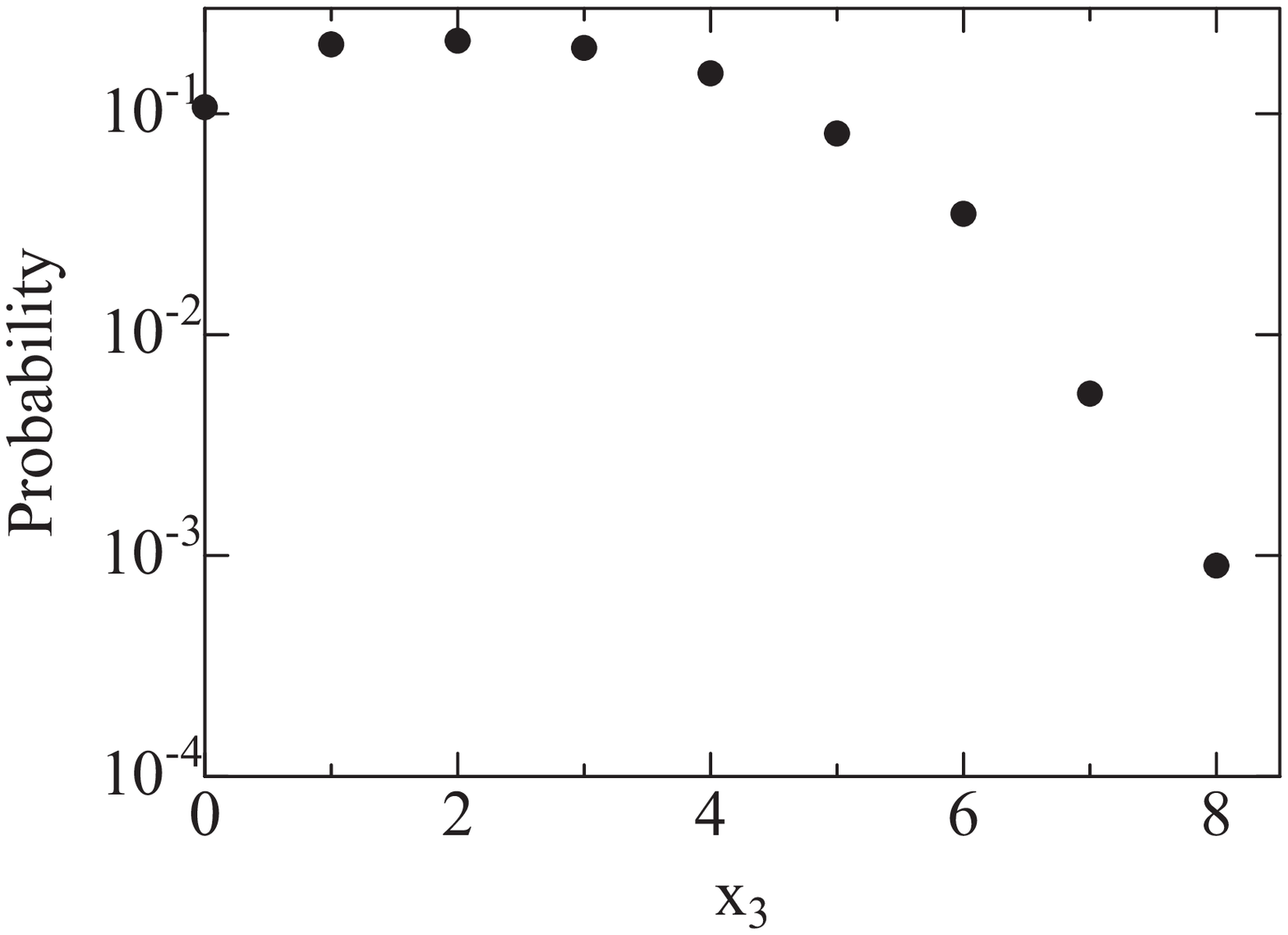}
\caption{Probability distribution of radiation after the third flight}
\end{figure}

\end{document}